\documentclass[12pt]{article}
\usepackage{natbib,graphicx}
\usepackage{amsmath}
\usepackage{amssymb}

\begin{document}

\title{Extended planetary chaotic zones}

\author{Ivan I. Shevchenko\/\thanks{E-mail:~i.shevchenko@spbu.ru} \\
Saint Petersburg State University, \\ 7/9 Universitetskaya
nab., 199034 Saint Petersburg, Russia \\
Institute of Applied Astronomy, Russian Academy of Sciences, \\
191187 Saint~Petersburg, Russia}

\date{}

\maketitle

\begin{abstract}
We consider the chaotic motion of low-mass bodies in two-body
high-order mean-motion resonances with planets in model planetary
systems, and analytically estimate the Lyapunov and diffusion
time\-scales of the motion in multiplets of interacting
subresonances corresponding to the mean-motion resonances. We show
that the densely distributed (though not overlapping) high-order
mean-motion resonances, when certain conditions on the planetary
system parameters are satisfied, may produce extended planetary
chaotic zones --- ``zones of weak chaotization,'' --- much broader
than the well-known planetary connected chaotic zone, the Wisdom
gap. This extended planetary chaotic zone covers the orbital range
between the 2/1 and 1/1 resonances with the planet. On the other
hand, the orbital space inner (closer to the host star) with
respect to the 2/1 resonance location is essentially long-term
stable. This difference arises because the adiabaticity parameter
of subresonance multiplets specifically depends on the particle's
orbit size. The revealed effect may control the structure of
planetesimal disks in planetary systems: the orbital zone between
the 2/1 and 1/1 resonances with a planet should be normally free
from low-mass material (only that occasionally captured in the
first-order 3/2 or 4/3 resonances may survive); whereas any
low-mass population inner to the 2/1 resonance location should be
normally long-lived (if not perturbed by secular resonances, which
we do not consider in this study).


\end{abstract}

\section{Introduction}
\label{intro}

The orbital architectures of planetary systems are established as
results of complex cosmogonical and dynamical processes, which
include planetary formation, close encounters, scattering,
migration in gas-dust disks. On the other hand, selection effects
favor discoveries of long-term stable systems. That is why
applications of stability criteria are necessary for explaining
the observed multitude of architectures of exoplanetary systems.

In this article, we reveal the existence of a weakly unstable
multi-resonant zone of dominant perturbative planetary influence,
which we call the {\it extended planetary chaotic zone} (EPCZ). It
covers the orbital range between the 2/1 and 1/1 mean-motion
resonances with the planet. Though weak, it is by far more
unstable (in a number of senses) than the orbital zone with
smaller orbital periods, inner to the 2/1 resonance. We argue that
observed structural patterns of planetesimal disks, such as the
2/1 resonance cut-off, may arise due to this effect.

The article is organized as follows. In Section~\ref{sec_Lem}, we
briefly review relevant theoretical issues on chaotic resonance
multiplets (including the standard map theory) in Hamiltonian
dynamics, concentrating on how the Lyapunov exponents and
diffusion rates in resonance multiplets can be estimated
analytically. In Section~\ref{sec_cmmr}, we apply the theory to
characterize the chaotic mean-motion resonances in planetary
systems. In Sections~\ref{sec_Ftr}, \ref{sec_SsSJ}, and
\ref{sec_msssE}, we characterize the resonances massively,
building the Farey trees of resonances in model planetary systems.
In Section~\ref{sec_cfc}, we compute covering factors of dynamical
chaos, as a function of test particles' orbit size, in the same
model systems. Section~\ref{sec_c} is devoted to general
discussions and conclusions.

\section{Lyapunov and diffusion timescales in resonance multiplets}
\label{sec_Lem}

We adopt a model of perturbed nonlinear resonance given by the
paradigmatic Hamiltonian
\begin{equation}
H = {{{\cal G} p^2} \over 2} - {\cal F} \cos \phi +
    a \cos(\phi - \tau) + b \cos(\phi + \tau)
\label{h}
\end{equation}
\noindent \citep{S99CM,S00JETP}, where the first two terms in
equation~(\ref{h}) represent the Hamiltonian $H_0$ of the
unperturbed pendulum, and periodic perturbations are represented
by the last two terms; $\phi$ is the pendulum angle (the resonance
phase angle), $p$ is the momentum, $\tau$ is the phase angle of
perturbation, $\tau = \Omega t + \tau_0$, and $\Omega$ is the
perturbation frequency, $\tau_0$ is the initial phase of the
perturbation; ${\cal F}$, ${\cal G}$, $a$, $b$ are constants.

Many resonant systems in physics and astronomy can be canonically
transformed to the perturbed pendulum model, which is thus
considered in some sense as a ``universal'' or ``first
fundamental'' model of nonlinear resonance; see \cite{C79,S20} for
details. In a generalized form, the perturbed pendulum model of
nonlinear resonance is given by Hamiltonian~(\ref{h}). In the next
section, we will see that this model perfectly corresponds to a
Hamiltonian description of high-order mean-motion resonances,
considered henceforth in this work.

In equation~(\ref{h}), the phase $\phi$ represents a linear
resonant combination of the angles of any original system; in the
next section, examples of such representation are given. The
momentum $p$ is proportional to the time derivative of $\phi$. The
momentum $p$ and phase $\phi$ form a pair of conjugated canonical
variables of the system defined by the Hamiltonian~(\ref{h}). The
system in non-autonomous, as the Hamiltonian explicitly depends on
time; thus, the system has one and a half degrees of freedom.

The small-amplitude libration frequency of the system on the
resonance modelled by the pendulum is $\omega_0 = ({\cal F
G})^{1/2}$. The ``adiabaticity parameter'', characterizing the
relative frequency of perturbation, is defined as
\begin{equation}
\lambda = {\Omega \over \omega_0}
\label{lambda}
\end{equation}
\noindent \citep{S20}.

The unperturbed resonance full width is defined as the maximum
distance (in momentum $p$) between its separatrices; it is equal
to $4 ({\cal F} / {\cal G})^{1/2}$. Therefore, the adiabaticity
parameter $\lambda$ characterizes the distance (in momentum $p$)
between the guiding and perturbing resonances in units of one
quarter of the full resonance width.

The rate of divergence of close trajectories (in the phase space
and in the logarithmic scale of distances) is characterized by the
maximum Lyapunov exponent. If the maximum Lyapunov exponent is
greater than zero, then the motion is chaotic \citep{C79,LL92}.
The inverse of this quantity, $T_{\rm L} \equiv L^{-1}$, is the
Lyapunov time. It represents the characteristic time of
predictable dynamics.

On the other hand, knowledge of {\it diffusion timescales} allows
one to judge on the possibility for effective transport in
action-like variables in the phase space of motion.

The Hamiltonian~(\ref{h}) includes three trigonometric terms,
corresponding to three interacting resonances, forming a resonance
triplet. Let the number of resonances in a resonance multiplet be
greater than three. In the case of non-adiabatic chaos ($\lambda
\gtrsim 1/2$) one may still apply, as an approximation, the theory
developed in \cite{S14PLA} for the triplet case, because the
influence of the ``far-away'' resonances is exponentially small
with $\lambda$. However, if chaos is adiabatic ($\lambda \lesssim
1/2$), the triplet approximation is no more valid for the
multiplet comprising more than three resonances. Therefore, let us
consider a limit case, namely, the case of infinitely many
interacting equal-sized and equally spaced resonances. The
standard map
\begin{eqnarray}
y_{i+1} &=& y_i + K \sin x_i \ \ \ (\mbox{mod } 2 \pi), \nonumber \\
x_{i+1} &=& x_i + y_{i+1} \ \ \ (\mbox{mod } 2 \pi) , \label{stm2}
\end{eqnarray}
\noindent describes the motion in just this case, as it is clear
from its Hamiltonian:
\begin{equation}
H = \frac{y^2}{2} + k \sum_{n=-M}^M \cos(x - n t) \label{h_stm2}
\end{equation}
\noindent \citep{C79}, where $M = \infty$, $k = K / (2 \pi)^2$.
The variables $x_i$, $y_i$ of the map~(\ref{stm2}) correspond to
the variables $x(t_i)$, $y(t_i)$ of the continuous
system~(\ref{h_stm2}) taken stroboscopically at time moduli $2
\pi$ (see, e.\,g., \citealt{C79}). Alongside with $\lambda$, the
standard map stochasticity parameter $K$ can be as well regarded
as a measure of resonance overlap, because the adiabaticity
parameter for the standard map is $\lambda \equiv \Omega/\omega_0
= 2 \pi K^{-1/2}$ \citep{S20}.

A formula for the maximum Lyapunov exponent of the standard map at
$K$ much greater than the critical value $K_\mathrm{G} =
0.9716...$, i.e., at $K \gg 1$, was derived by \cite{C79}:
\begin{equation}
L_\mathrm{st} \approx \ln
\frac{K}{2} . \label{L_st}
\end{equation}
\noindent Already at $K=6$ the difference between the theoretical
and numerical-experimental values of $L$ becomes less than
$\approx 2\%$ \citep{C79}.

The mapping period $T_\mathrm{map}$ of the map~(\ref{stm2}) can be
expressed in original time units of the system~(\ref{h_stm2}).
Therefore, for the maximum Lyapunov exponent of
system~(\ref{h_stm2}), at $K \gg 1$ (or, at $\lambda \ll 1$), one
has
\begin{equation}
L \approx \frac{L_\mathrm{st}}{T_\mathrm{map}} , \label{L_mp}
\end{equation}
where time is expressed in original time units. For the whole
interval of definition of $K$, we use the approximate numerical
and theoretical $L(K)$ functions obtained in
\cite{S04JETPL,S04PLA} for the main chaotic domain of the standard
map phase space at any $K$:
\begin{equation}
L(K) = \frac{1}{T_\mathrm{map}}
\cdot
\begin{cases}
0.1333 K , & \text{if $K < 1.1$}, \\
0.469 (K - 1.037)^{1/2} , & \text{if $1.1 \leq K < 4.4$}, \\
\ln \frac{K}{2} + \frac{1}{K^2} , & \text{if $K \geq 4.4$} ,
\end{cases}
\label{Lstmla}
\end{equation}
\noindent where
\begin{equation}
K = (2 \pi /\lambda)^2 ,
\label{Kla}
\end{equation}
\noindent and $T_\mathrm{map}$ is the perturbation period.

\begin{figure}[h!]
\centering
\includegraphics[width=0.8\textwidth]{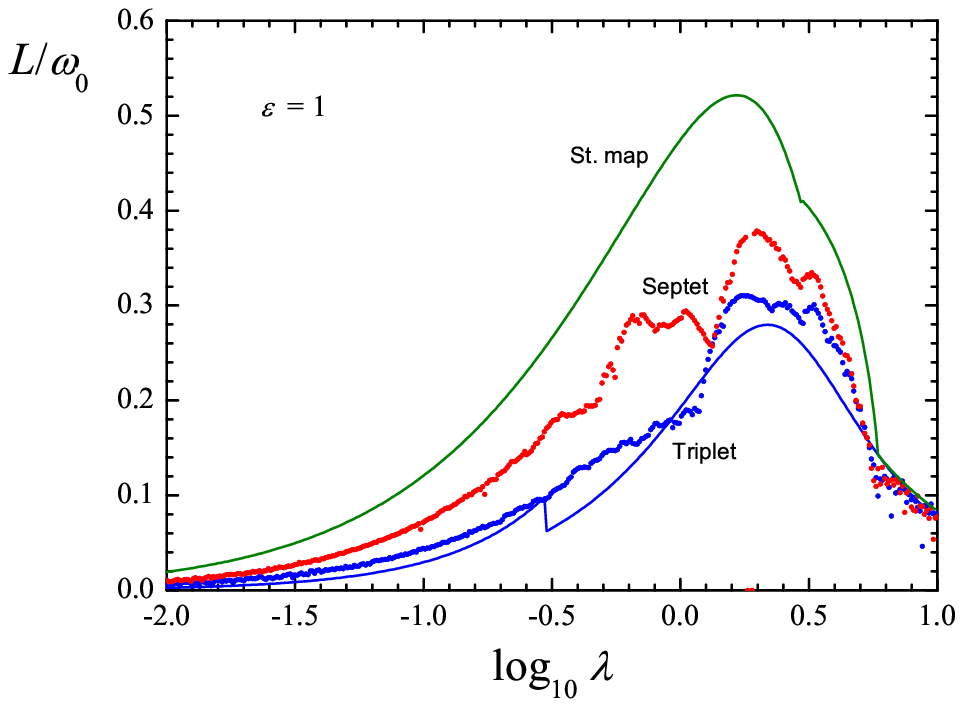} \\
\caption{The $\lambda$ dependences of the maximum Lyapunov
exponent (normalized by $\omega_0$) in resonance multiplets. Dots:
numerical-experimental data of \cite{S14PLA}. Green (upper) solid
curve: the standard map theory (given by equations~(\ref{Lstmla}))
for the ``infinitet''. Blue (lower) curve: the separatrix map
theory for the equally-spaced equally-sized triplet, as described
in \cite{S14PLA}. Adapted from \cite[Fig.~7]{S14PLA}.}
\label{L_mult}
\end{figure}

The $\lambda$ dependences, both numerical and theoretical, of the
maximum Lyapunov exponent (normalized by $\omega_0$) in multiplets
of equal-sized and equally spaced resonances are shown in
Fig.~\ref{L_mult}. Note that the normalization by $\omega_0$ means
that the given normalized maximum Lyapunov exponent is
adimensional.

In Fig.~\ref{L_mult}, the curve for the septet occupies an
intermediate position between the curve for the triplet and the
curve for the ``infinitet'', i.\,e., for the standard map.
Notwithstanding the large perturbation amplitude (resonances are
equal in size: in particular, for the triplet given by
equation~(\ref{h}) one has $\varepsilon \equiv a / {\cal F} = b /
{\cal F} = 1$), the numerical data for the triplet agrees well
with the separatrix map theory presented in \cite{S14PLA}; in
Fig.~\ref{L_mult}, this theory provides the lower solid curve.

The subresonances in the infinitet (the standard map case) start
to overlap, on decreasing $\lambda$, at $K_\mathrm{G} =
0.9716\ldots$ \citep{C79,Meiss92}, i.e., at $\lambda = 2 \pi
K_\mathrm{G}^{-1/2} \approx 6.37$. Therefore, the range in
$\lambda$ in Fig.~\ref{L_mult} almost totally corresponds to the
overlap conditions, except at $\log_{10} \lambda \gtrsim 0.8$.

The diffusion coefficient $D$ is defined as the mean-square spread
in a selected variable per time unit \citep{C79,Meiss92}; in the
standard map model, the selected variable is $y$. If $y$ is not
taken moduli $2 \pi$, then its variation is unbounded if $K >
K_\mathrm{G}$. At $K \gg 1$, the Lyapunov time $\approx \ln
\frac{K}{2}$ (Equation~(\ref{L_st})), and even adjacent iterated
values of the phase variables can be regarded as practically
independent. Then, the normal diffusion in $y$ at $K \gg 1$ has
the rate
\begin{equation}
D = \frac{\langle (\Delta y)^2 \rangle}{t} = \frac{1}{2} K^2
\label{Dql}
\end{equation}
\noindent \citep{C79}.

According to \cite{C79}, in the whole range $1 < K < \infty$, the
diffusion time, characterizing the mean time of transition between
neighbouring integer resonances, $T_\mathrm{d} \propto 1/D$, is
given by
\begin{equation}
T_\mathrm{d}^\mathrm{map} \simeq
\begin{cases}
100 \, (K - 1)^{-3} , & \text{if $1 < K \lesssim 4$}, \\
8 \pi^2 K^{-2} , & \text{if $K \gtrsim 4$},
\end{cases}
\label{NCh}
\end{equation}
\noindent which at $K \gg 1$ corresponds to the quasilinear
diffusion law~(\ref{Dql}). Generally, the standard map theory for
the diffusion rate is expected to be adequate if the number of
resonances in a considered resonance multiplet is large.

\section{Chaotic mean-motion resonances}
\label{sec_cmmr}

In the vicinities of high-order mean-motion two-body and
three-body resonances, the equations of motion are approximately
reducible to those of pendulum with periodic perturbations, given
by the Hamiltonian~(\ref{h}); see \cite{S20}. This reduction
provides an opportunity to analytically estimate the Lyapunov and
diffusion timescales of the motion, as described in the previous
Section.

Let us consider the restricted elliptical planar three-body
problem, with a passively gravitating test particle orbiting
around a primary mass $m_1$ and perturbed by the secondary mass
$m_2 < m_1$. In the vicinity of a mean-motion resonance $(k+q)/k$
(where $k \geq 1$ and $q \geq 1$ are integers) with the
gravitating binary, the Hamiltonian of the particle's barycentric
motion can be approximately represented as
\begin{equation}
H = {1\over2} \beta \Lambda^2 - \sum_{p=0}^{q} \varphi_{k+q, k+p,
k}\cos(\psi + p \varpi) \approx {1\over2} \beta \Lambda^2 -
\sum_{p=0}^{q} \varphi_{k+q, k+p, k}\cos(\psi - p \Omega t)
\label{HM}
\end{equation}
\noindent \citep{HM96,MH97}, where $\beta={3k^2} / a^2$, $\Lambda
= \Psi - \Psi_\mathrm{res}$, $\Psi = \left[ (1-\mu) a
\right]^{1/2}/k$, $\Psi_\mathrm{res} = \left[ (1-\mu)^2/(k^2 (k+q)
n_\mathrm{b}) \right]^{1/3}$, $\mu = m_2/(m_1 + m_2)$; $\varpi$ is
the longitude of the tertiary's (particle's) pericentre; $a$ and
$e$ are the tertiary's semimajor axis and eccentricity; $t$ is
time, and the frequency $\Omega$ is defined below in
equation~(\ref{Om}). The angle $\psi \equiv k l - (k+q)
l_\mathrm{b}$, where $l$ and $l_\mathrm{b}$ are the mean
longitudes of the tertiary and the primary binary.

The units are chosen in such a way that the total mass ($m_1+m_2$)
of the primary binary, the gravitational constant, the primary
binary's semimajor axis $a_\mathrm{b}$ are all equal to one. The
binary's mean longitude $l_\mathrm{b} = n_\mathrm{b} t$, and its
mean motion $n_\mathrm{b} = 1$, i.\,e., the time unit equals the
${1 \over 2 \pi}$th part of the binary's orbital period. The
binary's period $P_\mathrm{b}$ is set to $2 \pi$; its mean motion
$n_\mathrm{b} = 1$ and semimajor axis $a_\mathrm{b} = 1$.

The momentum $\Lambda$ and phase $\psi$ form a pair of conjugated
canonical variables of the system defined by the
Hamiltonian~(\ref{HM}); they can be put in correspondence to the
momentum $p$ and phase $\phi$ of the ``first fundamental'' model
Hamiltonian~(\ref{h}). We see that the model~(\ref{h}) can be put
in correspondence to the Hamiltonian description~(\ref{HM}) of
high-order mean-motion resonances, considered henceforth.

The system~(\ref{HM}) concerns the case of the outer (with respect
to the particle) perturber: the tertiary (the test particle)
orbits inside the secondary's (the perturber's) orbit around the
primary.

Note that the d'Alembert rule concerning the zero sum of integer
coefficients in the resonant angles is of course satisfied, but
indirectly, because the secondary's constant longitude of
pericentre is set equal to zero (on the d'Alembert rules see
\citealt{Morbi02}).

The integer non-negative numbers $k$ and $q$ define the resonance:
the ratio $(k+q)/k$ equals the ratio of the mean motions of the
tertiary (the particle) and the secondary (the perturber) in exact
resonance.

As described by equation~(\ref{HM}), if the perturber's orbit is
eccentric ($e_\mathrm{b} > 0$), then the resonance $(k+q)/k$
splits into a cluster of $q+1$ subresonances $p = 0, 1, \ldots,
q$, whose resonant arguments are given by the formula
\begin{equation}
\phi = \psi + p \varpi = k l - (q+k) l_\mathrm{b} + p \varpi .
\label{res_arg}
\end{equation}
The coefficients of the resonant terms, derived in \cite{HM96},
are given by
\begin{equation}
|\varphi_{k+q, k+p, k}|
\approx {\mu \over {2^q q \pi a_\mathrm{b}}} {q \choose p}
\epsilon^p \epsilon_\mathrm{b}^{q-p} , \label{coeff}
\end{equation}
\noindent where $\epsilon = {{e a_\mathrm{b}} /
{|a-a_\mathrm{b}|}}$, $\epsilon_\mathrm{b} = {{e_\mathrm{b}
a_\mathrm{b}} / {|a-a_\mathrm{b}|}}$, and ${i \choose j}$ is the
binomial coefficient. The approximation~(\ref{coeff}) is
applicable, if $\epsilon q < 1$ \citep{HM96}. Besides,
model~(\ref{HM}) is restricted to the resonances with $q \geq 2$.

The signs of the coefficients $\varphi_{k+q, k+p, k}$ alternate when
$p$ is incremented. Therefore, the coefficients with indices $p$
and $p+2$ always have the same sign. This means that at any choice
of the guiding resonance, its closest neighbours in the multiplet
have coefficients with equal signs.

In the model~(\ref{HM}), the coefficients $\varphi_{k+q, k+p, k}$ are
treated as constants. According to \cite{HM96} and \cite{MH97},
the frequencies $\omega_0$ of small-amplitude librations on
subresonances are given by
\begin{equation}
\omega_0 = (\beta | \varphi_{k+q, k+p,k} |)^{1/2} \approx
\frac{n_\mathrm{b}}{1-\alpha} \left[ (1-\mu)
\mu {{2^{2-q} q} \over {3\pi}} {q \choose p}
\alpha \epsilon^p
\epsilon_\mathrm{b}^{q-p} \right]^{1/2} , \label{om0}
\end{equation}
\noindent and the perturbation frequency is
\begin{equation}
\Omega \equiv - \dot \varpi \approx \frac{\alpha
b^{(1)}_{3/2}(\alpha)}{8 \left[ (1-\mu) a_\mathrm{b} \right]^{1/2} } \approx
\frac{(1-\mu) \mu}{2 \pi} n_\mathrm{b}
\frac{\alpha^{1/2}}{(1-\alpha)^2} ,
\label{Om}
\end{equation}
\noindent where $b^{(1)}_{3/2}(\alpha)$ is a Laplace coefficient,
and
\begin{equation}
\alpha \equiv a/a_\mathrm{b} = \left[ k/(k+q) \right]^{2/3} .
\end{equation}

Following \cite{MH97}, for the effective stochasticity parameter
$K$ in the subresonance multiplet we take
\begin{equation}
K_\mathrm{eff} = \beta \psi_\mathrm{mid, 0}
\left( \frac{\pi}{\mu A} \right)^2 ,
\label{Keff}
\end{equation}
\noindent where
\begin{equation}
\beta = \frac{3 k^2}{{a_\mathrm{b}}^2} \simeq \frac{4 q^2}{3
\epsilon^2} , \label{beta1}
\end{equation}
and
\begin{equation}
A = \frac{\alpha b^{(1)}_{3/2}(\alpha)}{8 ( (1-\mu)
a_\mathrm{b} )^{1/2} } \simeq \frac{1}{4 \pi} \frac{ \alpha^{1/2}
}{\epsilon^2} . \label{Ac1}
\end{equation}
\noindent \noindent Further on, for evaluating $\beta$ and $A$, we
use the non-approximated expressions, i.e., the first ones in
equations (\ref{beta1}) and (\ref{Ac1}). The stochasticity
parameter $K$ of the standard map theory \citep{C79} is the same,
in its dynamical sense, as the given $K_\mathrm{eff}$; otherwise,
in the standard map case, the resonance multiplet is infinite.

For the guiding subresonance we take the strongest one, that in
the middle of the multiplet. As soon as $p = 0, 1, 2, 3, ..., q$,
the ``middle'' value of $p$ is $p_\mathrm{mid} = (q + 1)/2 - 1$;
and, if $q=1$, then we take $p_\mathrm{mid} = 1$.

Let us calculate the width of the chaotic multiplet (that with
overlapping subresonances, $K>K_\mathrm{G}$). First of all, we
define technical quantities. For the first subresonance ($p=0$),
such quantity is
\begin{equation}
\psi_0 = \frac{\mu}{\pi q} {q \choose 0} \left( \frac{\epsilon_\mathrm{b}}{2}
\right)^q =
\frac{\mu}{\pi q} \left( \frac{\epsilon_\mathrm{b}}{2} \right)^q ,
\end{equation}
\noindent whereas, for the last one ($p=q$),
\begin{equation}
\psi_q = \frac{\mu}{\pi q} {q \choose q}
\left( \frac{\epsilon_0}{2} \right)^q
= \frac{\mu}{\pi q} \left( \frac{\epsilon_0}{2} \right)^q ,
\end{equation}
\noindent and, for the middle one,
\begin{equation}
\psi_\mathrm{mid, 0} = \frac{\mu}{2^q \pi q}
{q \choose p_0} \epsilon_0^{p_0}
\epsilon_\mathrm{b}^{q-p_0}
\end{equation}
\noindent at $\epsilon = \epsilon_0$, and
\begin{equation}
\psi_{\mathrm{mid, max}} = \frac{\mu}{2^q \pi q}
{q \choose p_0}
\epsilon_\mathrm{max}^{p_0}
\epsilon_\mathrm{b}^{q-p_0}
\end{equation}
\noindent at $\epsilon = \epsilon_\mathrm{max}$.

The distance, in the canonical momentum variable, between the
subresonances is
\begin{equation}
\delta \Lambda = 2 \mu A/\beta ;
\end{equation}
\noindent see \cite[equation~(28)]{MH97}. For the first and last
subresonances in the multiplet, the half-widths are given by
\begin{equation}
\Delta \Lambda_0 = 2 ( \psi_0/\beta )^{1/2}, \quad \Delta \Lambda_q =
2 ( \psi_q/\beta )^{1/2} ,
\end{equation}
\noindent and, summing, for the total width $\Delta a_\mathrm{ch}$
of the subresonance multiplet one has
\begin{equation}
\Delta a_\mathrm{ch} =
\frac{2 k a_\mathrm{b}^{1/2}}{ (1-\mu)^{1/2} }
\left( q \delta \Lambda + \Delta \Lambda_0 +
\Delta \Lambda_q \right) .
\end{equation}
\noindent
If $\Delta a_\mathrm{ch} < 2 \Delta \Lambda_0$, we take $\Delta a_\mathrm{ch}
= 2 \Delta a_0$, and if $\Delta a_\mathrm{ch} < 2 \Delta
\Lambda_q$, we take $\Delta a_\mathrm{ch} = 2 \Delta a_q$.

Note that the total width of the chaotic multiplet is calculated
here taking into account the half-widths of the boundary
subresonances, as bounded by their {\it unperturbed} separatrices.
The widths of the perturbed (splitted) separatrices can also be
calculated (see \citealt{S08PLA,S20}), but we ignore them in view
of the dominating widths of the considered multiplets themselves.

To apply in the next sections, let us write down also an
expression for the half-width $\Delta a_\mathrm{cr}$ of the Wisdom
gap (the planetary connected chaotic zone). In units of the
perturber's semimajor axis $a_\mathrm{b}$, it is given by
\begin{equation}
\Delta a_\mathrm{cr}/a_\mathrm{b} \approx 1.6 \mu^{2/7}
\label{Wgw}
\end{equation}
\noindent \citep{DQT89,MD99}; concerning the accuracy of the
numerical coefficient, see discussion in \cite{S20}.

Consider now the diffusion rates. As follows from
equation~(\ref{Dql}), the diffusion coefficients are given by
\begin{equation}
D(I_0) = \frac{\pi p_0^2 \psi_\mathrm{mid, 0}^2}{2 \mu A} ,
\label{DI0}
\end{equation}
\begin{equation}
D(I_\mathrm{max}) = \frac{\pi p_0^2 \psi_{\mathrm{mid, max}}^2}{2
\mu A} ,
\label{DImax}
\end{equation}
\noindent where $I_0 = e_0^2/2$, $I_\mathrm{max} =
e_\mathrm{max}^2/2$. To compute the removal time, we set
$e_\mathrm{max} = 0.4$, as this eccentricity value is normally
sufficient for reaching typical secular resonances in the inner Solar
system (see, e.g., \citealt{Morbi02}).

We do not take the particle's initial eccentricity equal to a
particular constant value (as was adopted in \citealt{HM96,MH97}),
but take it equal to the forced eccentricity. In the perturber's
vicinity, according to \cite[equation~(33)]{HP86I}, the latter is
given by
\begin{equation}
e_\mathrm{f,HP} = \frac{2.24 \, \mu}{(1-\alpha)^2} , \label{efHP}
\end{equation}
\noindent and, far from the perturber, according to
\cite[equation~(4)]{H78AA}, it is given by
\begin{equation}
e_\mathrm{f,H} = \frac{2}{\pi} \cdot \frac{2.5 \, \alpha
e_\mathrm{b}}{(1 - {e_\mathrm{b}}^2)} = \frac{5}{\pi} \cdot \frac{\alpha
e_\mathrm{b}}{(1 - {e_\mathrm{b}}^2)} .
\label{efH}
\end{equation}
\noindent This is the time-averaged quantity, hence the
coefficient $2/\pi$. At a given value of $a$, if $e_\mathrm{f,HP}
> e_\mathrm{f,H}$, then we take $e_0 = e_\mathrm{f,HP}$, else we take $e_0
= e_\mathrm{f,H}$.

In accord with equations~(\ref{DI0}) and (\ref{DImax}), the
diffusion time is therefore given by
\begin{equation}
T_\mathrm{d} =
\begin{cases}
\displaystyle{
\frac{p_{I_0} p_{I_\mathrm{max}}}{\left[ D(I_0) D(I_\mathrm{max}) \right]^{1/2}} ,
}
& \text{if $q > 2$}, \\
\qquad
\displaystyle{
\frac{I_\mathrm{max}^2}{D(I_0)} ,
}
& \text{if $q \leq 2$} .
\end{cases}
\label{Tr2Pi}
\end{equation}
\noindent To use the standard map theory, we set $K=K_\mathrm{eff}$.

In the standard map theory, the Lyapunov exponent $L$ is given by
formula~(\ref{Lstmla}). Therefore, the Lyapunov time for
Hamiltonian~(\ref{HM}), in the perturber's orbital periods, can be written
down as
\begin{equation}
T_\mathrm{L} = \frac{1}{2 \mu A L} .
\end{equation}
\noindent
The diffusion time $T_\mathrm{d}$ is given by formula~(\ref{Tr2Pi});
therefore, in the perturber's orbital periods, the removal time is
\begin{equation}
T_\mathrm{r} = \frac{1}{2 \pi} T_\mathrm{d} .
\end{equation}

\section{The Farey tree of mean-motion resonances}
\label{sec_Ftr}

The Farey tree technique is used in the number theory to organize
rational numbers \citep{HW79}. Here we use it to organize
mean-motion resonances in a clear and straightforward way.

The Farey tree is built as follows. Consider some rational numbers
$m'/n'$ and $m''/n''$ that are ``neighbouring'', i.e., $m' n'' -
m'' n' = 1$. Let them form the first level of the tree. Then, the
second level of the tree is formed by a ``mediant,'' given by the
formula $m'''/n''' = (m' + m'')/(n' + n'')$. Each next level is
formed by taking mediants of the numbers obtained at all preceding levels.
Thus, the third level comprises two mediants ($(m' + m''')/(n' +
n''')$ and $(m''' + m'')/(n''' + n'')$) of three numbers at two
lower levels, the fourth level comprises four mediants of five
numbers at three lower levels, and so on. If, at the first level,
one takes $m'/n'=0$ and $m''/n''=1$, then the Farey tree,
generated up to infinity, will comprise all rational numbers in
the [0,~1] closed interval. For details, see
\cite[pp.~814--815]{Meiss92}; a graphical scheme of the Farey tree
construction is given in fig.~26 in \cite{Meiss92}.

Concerning the motion inner to the perturber in our planetary
problem, the ratio of orbital frequencies (mean motions) of the
particle and the perturber is greater than one; therefore,
representing mean-motion resonances by rational numbers, we define
the resonances as reciprocals of the rational numbers in the Farey
tree generated in the [0,~1] segment.

Recall that the order of a mean-motion resonance is given by the
difference between the numerator and the denominator in its
rational-number representation. It is important that, at each
consequent level of the Farey tree, the order of any generated
mean-motion resonance may only rise or stay constant; indeed, for
the rational-number mediant $m'''/n''' = (m' + m'')/(n' + n'')$
the order of the corresponding resonance is $q''' = n' + n'' - m'
- m''$, i.e., it is equal to $q' + q''$, the sum of the orders of
two lower-level generating resonances. As soon as the orders are
non-negative, the generated resonance order cannot decrease. It is
also important to note that the Farey tree covers and organizes
the full set of rational numbers \citep{HW79,Meiss92};
accordingly, it covers and organizes the entire set of mean-motion
resonances.

For two generating integer resonances $p/1$ and $(p+1)/1$, the
mediant will be $(2p+1)/2$. Therefore, the half-integer resonances
are the mediants for the integer ones, and so on. The number of
all resonances up to level $k$ is $N_\mathrm{res} = 2^{k-1} + 1$.

For the first and second generating rational numbers at the first
level of the Farey tree, we take, respectively, 0/1 and 1/1. They
correspond to the mean-motion resonances 1/0 and 1/1 of the
particle with the perturber (in its turn, these two resonances
correspond to the test particle's semimajor axis $a=0$ and $a=1$,
in units of the perturber's semimajor axis). Then, following the
outlined above algorithm, we obtain the resonances 2/1 (at the
second level of the tree); 3/1 and 3/2 (at the third level); 4/1,
5/2, 5/3, 4/3 (at the fourth level); and so on.

\section{The ``Sun -- Jupiter -- minor body''  model system}
\label{sec_SsSJ}

Let us consider our Solar system with Jupiter regarded as a unique
perturber, i.e., we ignore all other planets. Therefore, in the
formulas of Section~\ref{sec_cmmr}, we set the mass parameter $\mu
= 1/1047$, the secondary's eccentricity $e_\mathrm{b} = 0.048$ and
its orbital period $P_\mathrm{b} = 11.86$~yr.

Using the algorithm described in Section~\ref{sec_Ftr}, we
generate mean-motion resonances in the inner Solar system up to
level~10 of the Farey tree, and compute the Lyapunov times,
removal times, and widths of the chaotic resonance multiplets,
using formulas given in Section~\ref{sec_cmmr}.

In Fig.~\ref{J_lgK_a}, we illustrate the mean-motion resonances in
the inner Solar system. In the left panel of this Figure, the
stochasticity parameter $K$ (blue dots) of the resonances is shown
as a function of the tertiary's semimajor axis $a$. The vertical
green line marks the location of the 2/1 resonance with Jupiter.
The horizontal magenta and blue dotted lines correspond to
$K=K_\mathrm{G}$ and $K=4$, respectively. One may see that in the
orbital range between the 2/1 and 1/1 resonances the values of $K$
are orders of magnitude greater than those in the range between
the 0/1 and 2/1 resonances. In the right panel, the same set of
resonances is displayed, but for the product $q \epsilon$ (blue
dots). The horizontal red line marks the $q \epsilon = 1$ limit.
We see that for the most of the high-order resonances, the product
$q \epsilon > 1$; this means that the adopted theory can be used
solely as an extrapolation.

At smaller values of $\mu$ and $e_\mathrm{b}$, one may use the
theory without any extrapolation, as we demonstrate in the next
Section.

In Fig.~\ref{J_TL_a}, left panel, the Lyapunov time $T_\mathrm{L}$
(olive dots) is shown as a function of the tertiary's semimajor
axis $a$. The vertical green line marks the location of the 2/1
resonance with Jupiter. The vertical red line marks the inner
border of the Wisdom gap (the planetary connected chaotic zone) of
Jupiter, and the double vertical black, light magenta, blue, and
magenta lines mark the borders of the Wisdom gaps of Mercury,
Venus, Earth, and Mars, respectively. The Wisdom gap borders'
locations are given by equation~(\ref{Wgw}).

In the right panel, the removal times $T_\mathrm{r}$ (olive dots)
are shown for the resonances with $K>K_\mathrm{G}$. The two panels
certify that, in the orbital range between the 2/1 and 1/1
resonances, the $T_\mathrm{L}$ and $T_\mathrm{r}$ values are
orders of magnitude less than those in the range between the 0/1
and 2/1 resonances.

In Fig.~\ref{J_Width_a}, the widths $\Delta a_\mathrm{ch}$ (in
tertiary's semimajor axis) of the subresonance multiplets of
mean-motion resonances are displayed (the blue dots). For the
resonances with $K<K_\mathrm{G}$, the widths are set to zero. The
vertical green line marks the location of the 2/1 resonance with
Jupiter. The vertical red line marks the inner border of the
Wisdom gap of Jupiter. We see that the total width of the chaotic
resonances to the left of the 2/1 resonance is just zero, in
contrast to the situation to the right of the 2/1 resonance, where
a lot of chaotic resonances of significant measure are present.

\section{The ``star -- super-Earth -- minor body'' model system}
\label{sec_msssE}

Now let us consider a system with a much smaller value of the mass
parameter $\mu$: a ``Solar-like star -- super-Earth'' system. Mass
of the model super-Earth is set equal to three Earth masses, i.e.,
$\mu = 10^{-5}$; and the super-Earth's orbit eccentricity
$e_\mathrm{b} = 0.005$. As in the previous Section, we generate
the mean-motion resonances in the inner model system up to
level~10 of the Farey tree, and calculate the Lyapunov and removal
times and the widths of the chaotic subresonance multiplets of the
mean-motion resonances.

In Fig.~\ref{SE_lgK_a}, we illustrate the mean-motion resonances
in our model exoplanet system. In the left panel of this Figure,
the stochasticity parameter $K$ of the chaotic subresonance
multiplets of the mean-motion resonances is shown as a function of
the tertiary's semimajor axis $a$ (blue dots). The semimajor axis
of the super-Earth orbit is set to one. The vertical green line
marks the location of the 2/1 resonance with the super-Earth. The
horizontal magenta and blue dotted lines correspond to
$K=K_\mathrm{G}$ and $K=4$, respectively. As in
Section~\ref{sec_SsSJ} above, one may see that in the range
between the 2/1 and 1/1 resonances the values of $K$ are typically
orders of magnitude greater than in the range between the 0/1 and
2/1 resonances. Also we display (in the right panel) the same
resonances, but for the product $q \epsilon$ (blue dots). The
horizontal red line marks the $q \epsilon = 1$ limit. We see that
for all resonances, the product $q \epsilon < 1$; this means that
the adopted theory is everywhere valid.

In Fig.~\ref{SE_TL_a}, left panel, the Lyapunov time
$T_\mathrm{L}$ (olive dots) is shown is shown as a function of the
tertiary's semimajor axis $a$. The vertical green line marks the
location of the 2/1 resonance with the super-Earth. The vertical
red line marks the inner border of the super-Earth's Wisdom gap.
In the right panel, the removal times $T_\mathrm{r}$ (olive dots)
are shown for the resonances with $K>K_\mathrm{G}$. As in
Section~\ref{sec_SsSJ} above, the panels of Fig.~\ref{SE_TL_a}
make it clear that, in the range between the 2/1 and 1/1
resonances, the $T_\mathrm{L}$ and $T_\mathrm{r}$ values are
orders of magnitude smaller than those in the range between the
0/1 and 2/1 resonances.

In Fig.~\ref{SE_Width_a}, the widths $\Delta a_\mathrm{ch}$ (blue
dots) of the subresonance multiplets of mean-motion resonances are
displayed. For the subresonance multiplets with $K<K_\mathrm{G}$,
the widths are set to zero. The vertical green line marks the
location of the 2/1 resonance with the super-Earth. The vertical
red line marks the inner border of the super-Earth's Wisdom gap.
As in Section~\ref{sec_SsSJ}, we see that the total width of
chaotic resonances to the left of the 2/1 line is zero, whereas to
the right of the 2/1 line there is a lot of broad chaotic
resonances.

\section{Covering factors of dynamical chaos}
\label{sec_cfc}

Let us define the covering factor of chaos as the sum of the
widths of the mean-motion resonances with $K>K_\mathrm{G}$ in a
particular range of the initial orbital radii of the test minor
body. This notion may seem similar to ``optical depths,'' used in
\cite{Q11} and \cite{HL18} to characterize resonance ensembles,
but there exists a qualitative difference: the covering factor, as
introduced here, concerns chaotic resonances (those with
overlapping subresonances), whereas the ``optical depths'' take
into account the widths of all resonances.

In the model Solar system, defined in Section~\ref{sec_SsSJ}, the
sole perturber is Jupiter and, therefore, the mass parameter $\mu
= 10^{-3}$. The covering factor of chaos in the inner Solar system
(apart from the Wisdom gap of Jupiter), for the resonance ensemble
up to the Farey tree level~10, is shown, as a function of
Jupiter's eccentricity $e_\mathrm{planet} = e_\mathrm{b}$, in
Fig.~\ref{W_ecc_mu-3}. The horizontal blue dotted line represents
the covering factor (the radial half-width) of the Wisdom gap of
Jupiter. We see that the covering factor of the chaotic resonances
inner (in orbital radius) to the Wisdom gap is always much less
than the covering factor (the relative radial size) of the Wisdom
gap itself.

In the model exoplanet system, as defined in
Section~\ref{sec_msssE}, the mass parameter $\mu = 10^{-5}$. The
covering factor of the mean-motion resonances with
$K>K_\mathrm{G}$ in the inner system up to the Farey tree level~10
is shown, as a function of the planet's eccentricity
$e_\mathrm{planet}$, in Fig.~\ref{W_ecc_mu-5}. The horizontal blue
dotted line represents the covering factor (the radial half-width)
of the super-Earth's Wisdom gap. We see that, with increasing the
perturber's eccentricity, the covering factor of the chaotic
resonances inner (in orbital radius) to the perturber's Wisdom gap
starts to dominate over the covering factor of the Wisdom gap
rather rapidly.

\section{Discussion and conclusions}
\label{sec_c}

As shown above in Sections \ref{sec_SsSJ} and \ref{sec_msssE}, the
densely distributed (though not overlapping) high-order
resonances, when certain conditions for the planetary system
parameters are satisfied, may produce an extended planetary
chaotic zone (EPCZ) --- the weak instability zone, disconnectedly
but densely extending, in orbital radius, the planetary chaotic
zone down to the 2/1 resonance with the planet. Therefore, the
extended planetary chaotic zone covers the orbital range between
the 2/1 and 1/1 resonances with the planet. On the other hand, the
orbital space inner to the 2/1 resonance can be essentially
long-term stable.

What is the cause of this difference? Let us demonstrate that it
can be qualitatively explained as due to a specific behavior of
the dependence of the adiabaticity parameter of subresonance
multiplets on the value of the particle orbit's semimajor axis.
The adiabaticity parameter is given by equation~(\ref{lambda}):
$\lambda = \Omega/\omega_0$. Using equations~(\ref{om0}) and
(\ref{Om}) for the frequencies $\omega_0$ and $\Omega$, one
arrives at
\begin{equation}
\lambda = {\Omega \over \omega_0} = C \cdot
(1-\alpha)^{\frac{q}{2}-1}, \label{lambda1}
\end{equation}
\noindent where $\alpha = a/a_\mathrm{b}$, as defined above, and
the coefficient
$C=C(\mu, q, e, e_\mathrm{b})$ does not depend on $\alpha$. One
may see that, at $q>2$, $\lambda$ tends to zero, if $\alpha \to
1$, and it tends to $C$, if $\alpha \to 0$.

Equation~(\ref{lambda1}) is valid if the particle's proper
eccentricity is large enough, namely, if it is much greater than
the forced eccentricity: $e_\mathrm{proper} \gg
e_\mathrm{forced}$. As mentioned above, the forced eccentricity in
immediate vicinities of the perturber is given, according to
\cite{HP86I}, by equation~(\ref{efHP}), and, far from the
perturber it is given, according to \cite{H78AA}, by
equation~(\ref{efH}).

If $e_\mathrm{proper} \ll e_\mathrm{forced}$, one should
substitute $e = e_\mathrm{forced}$ in the expression for
$\omega_0$, equation~(\ref{om0}). Then, for the adiabaticity
parameter, at locations close to the perturber, one has
\begin{equation}
\lambda = C' \cdot (1-\alpha)^{q-1}
\label{lambda2}
\end{equation}
\noindent with the coefficient $C'=C'(\mu, q, e_\mathrm{b})$. At
locations far from the perturber, one has
\begin{equation}
\lambda = C'' \cdot (1-\alpha)^{\frac{q}{2}-1} \cdot
\alpha^{-\frac{q}{4}} \label{lambda3}
\end{equation}
\noindent with $C''=C''(\mu, q, e_\mathrm{b})$.
Equations~(\ref{lambda2}) and (\ref{lambda3}) demonstrate that in
the whole range $0 < \alpha < 1$, at $q>1$, the adiabaticity
parameter $\lambda$ increases if $\alpha$ is decreased. In other
words, $\lambda$ becomes larger for orbits more distant from the
perturber and more closer to the host star.

Thus, for high-order mean-motion resonances, either at
$e_\mathrm{proper} \gg e_\mathrm{forced}$ or at $e_\mathrm{proper}
\ll e_\mathrm{forced}$, the adiabaticity parameter $\lambda$
increases if $\alpha$ is decreased from 1 down to 0.

\begin{figure}[h!]
\centering
\begin{tabular}{cc}
\includegraphics[width=0.5\textwidth]{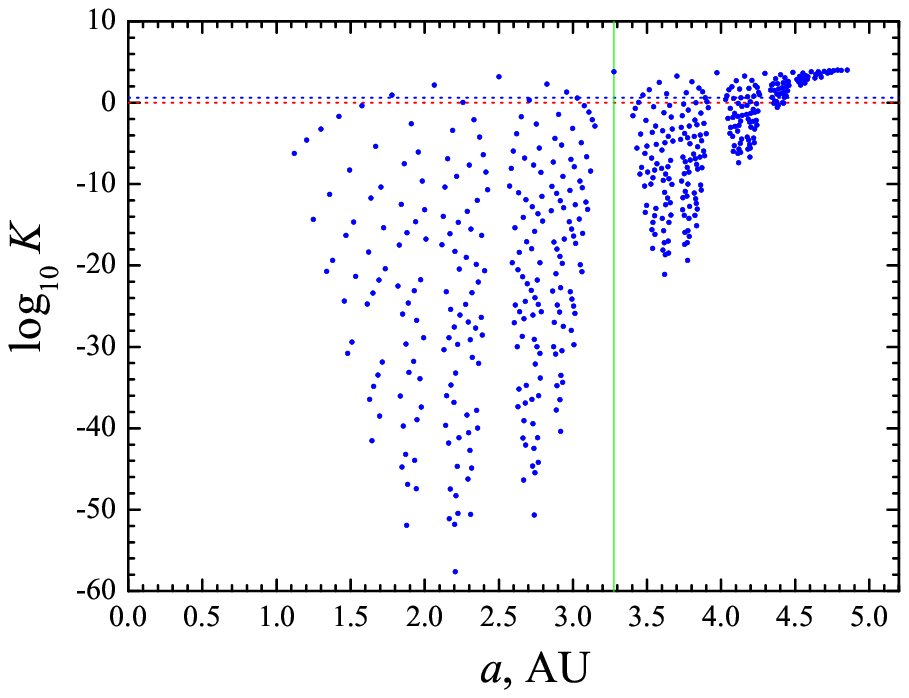}
&
\includegraphics[width=0.49\textwidth]{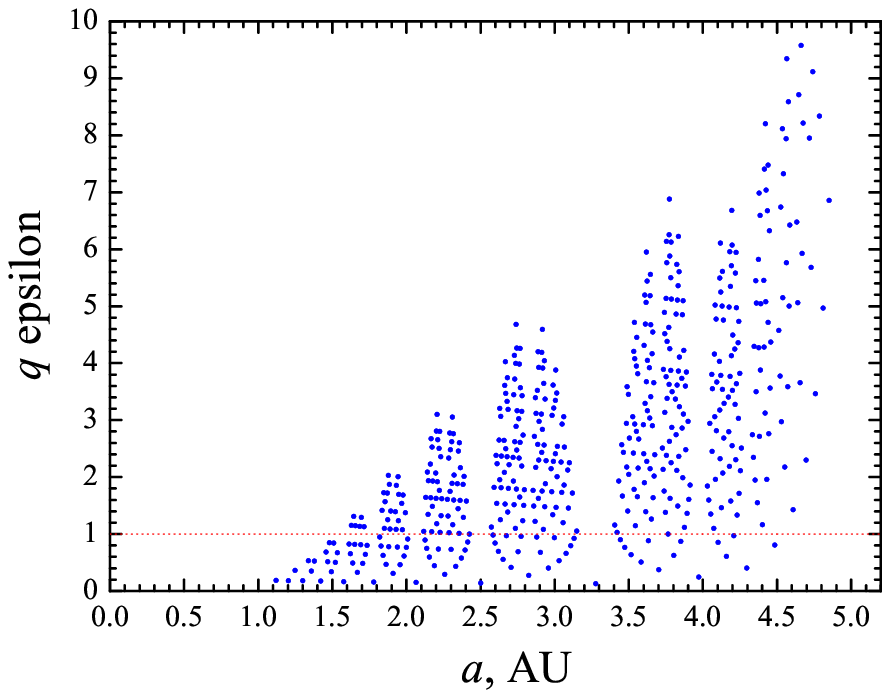}
\end{tabular}
\caption{The mean-motion resonances in the inner Solar system up
to the Farey tree level~10. Left panel: the stochasticity
parameter $K$ (blue dots) as a function of the tertiary's
semimajor axis $a$. Vertical green line: the 2/1 resonance with
Jupiter. Magenta and blue dotted horizontal lines: the
$K=K_\mathrm{G}$ and $K=4$ limits, respectively. Right panel: the
product $q \epsilon$ (blue dots) as a function of $a$; red
horizontal line: the $q \epsilon = 1$ limit.} \label{J_lgK_a}
\end{figure}

\begin{figure}[h!]
\centering
\begin{tabular}{cc}
\includegraphics[width=0.5\textwidth]{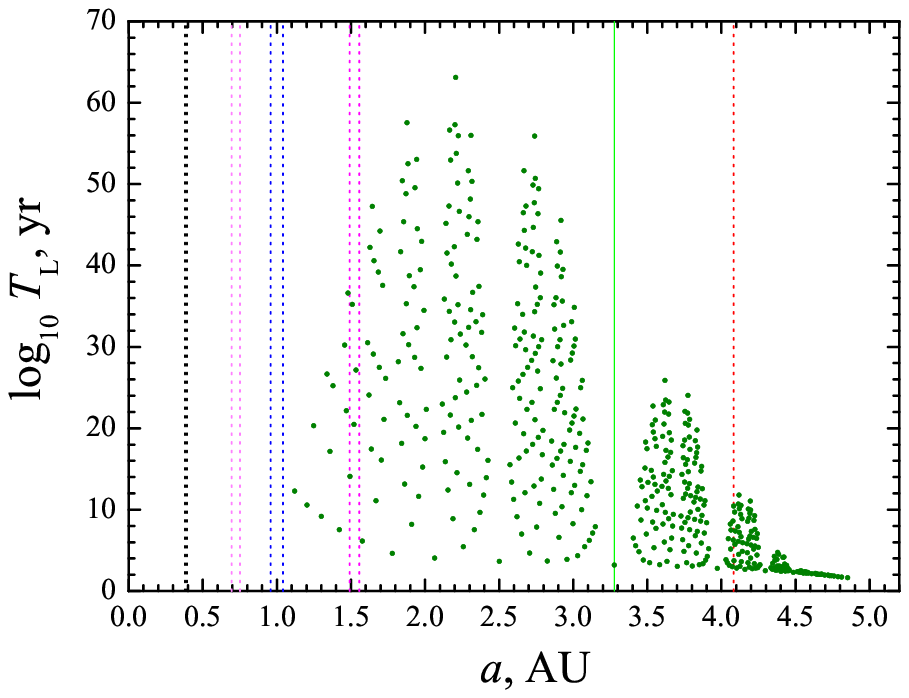}
&
\includegraphics[width=0.5\textwidth]{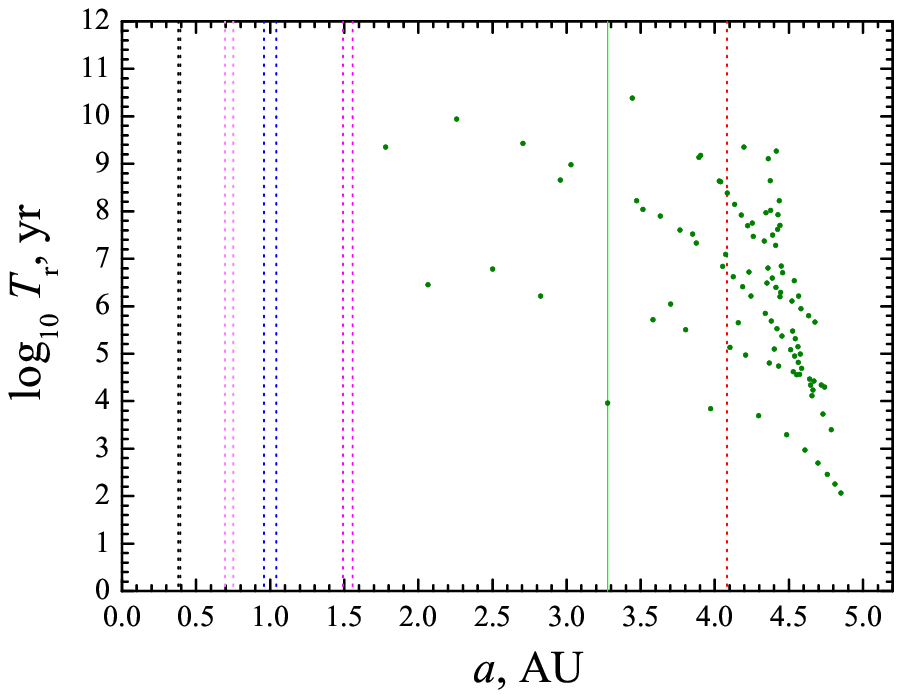}
\end{tabular}
\caption{The Lyapunov and removal times, in the same model as in
Fig.~\ref{J_lgK_a}. Left panel: the Lyapunov time $T_\mathrm{L}$
(olive dots) as a function of the tertiary's semimajor axis $a$.
Vertical green line: the 2/1 resonance with Jupiter. Red vertical
line: the inner border of the Wisdom gap of Jupiter. Double
vertical black, light magenta, blue, and magenta lines: the
borders of the Wisdom gaps of Mercury, Venus, Earth, and Mars,
respectively. Right panel: the removal time $T_\mathrm{r}$ (olive
dots) as a function of semimajor axis $a$. Solely the resonances
with $K>K_\mathrm{G}$ are presented.} \label{J_TL_a}
\end{figure}

\begin{figure}[h!]
\centering
\includegraphics[width=0.8\textwidth]{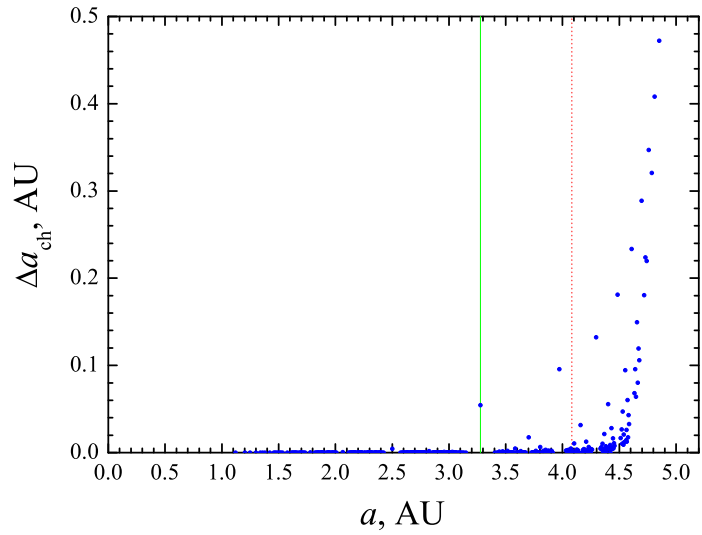} \\
\caption{The widths of subresonance multiplets of resonances, in
the same model as in Figs~\ref{J_lgK_a} and \ref{J_TL_a}. Blue
dots: the widths $\Delta a_\mathrm{ch}$ of subresonance
multiplets. For the resonances with $K < K_\mathrm{G}$ the widths
are set to zero. Green vertical line: the 2/1 resonance with
Jupiter. Red vertical line: the inner border of the Wisdom gap of
Jupiter.} \label{J_Width_a}
\end{figure}

\begin{figure}[h!]
\centering
\begin{tabular}{cc}
\includegraphics[width=0.5\textwidth]{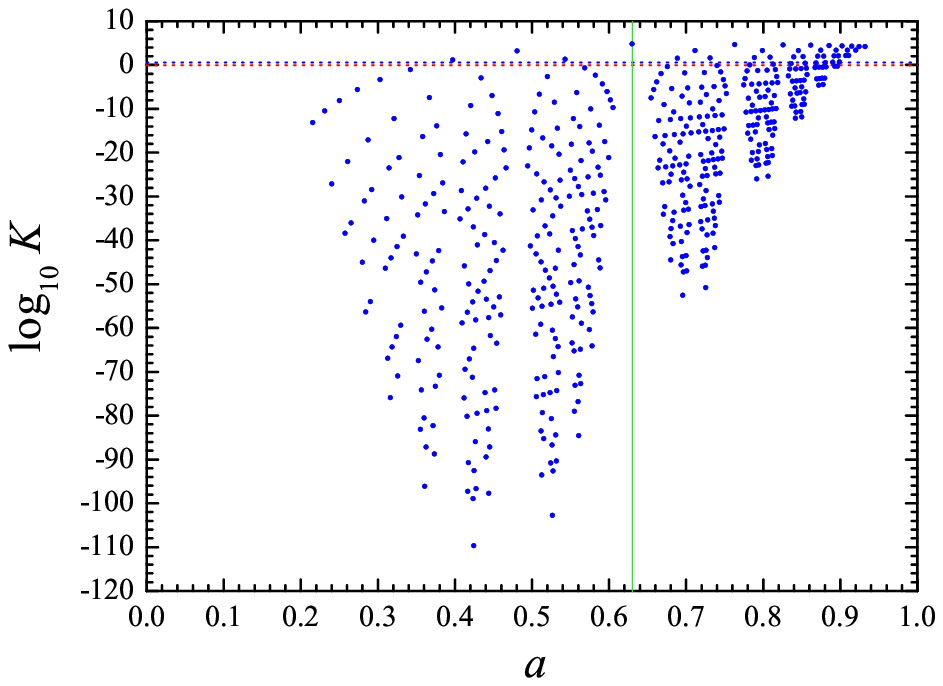}
&
\includegraphics[width=0.49\textwidth]{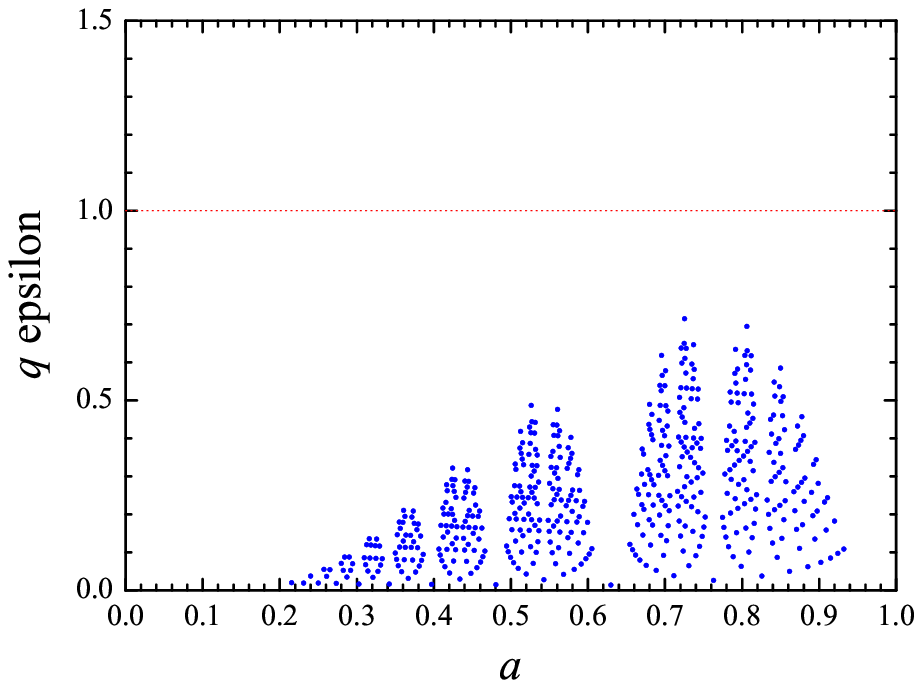}
\end{tabular}
\caption{The mean-motion resonances in the model ``Sun-like star
-- super-Earth -- minor body'' system up to the Farey tree
level~10. Left panel: the stochasticity parameter $K$ (blue dots)
as a function of the tertiary's semimajor axis $a$. The semimajor
axis of the super-Earth is set to one. Green vertical line: the
2/1 resonance with the super-Earth. Magenta and blue dotted
horizontal lines: the $K=K_\mathrm{G}$ and $K=4$ limits,
respectively. Right panel: the product $q \epsilon$ (blue dots) as
a function of $a$; red horizontal line: the $q \epsilon = 1$
limit.} \label{SE_lgK_a}
\end{figure}

\begin{figure}[h!]
\centering
\begin{tabular}{cc}
\includegraphics[width=0.5\textwidth]{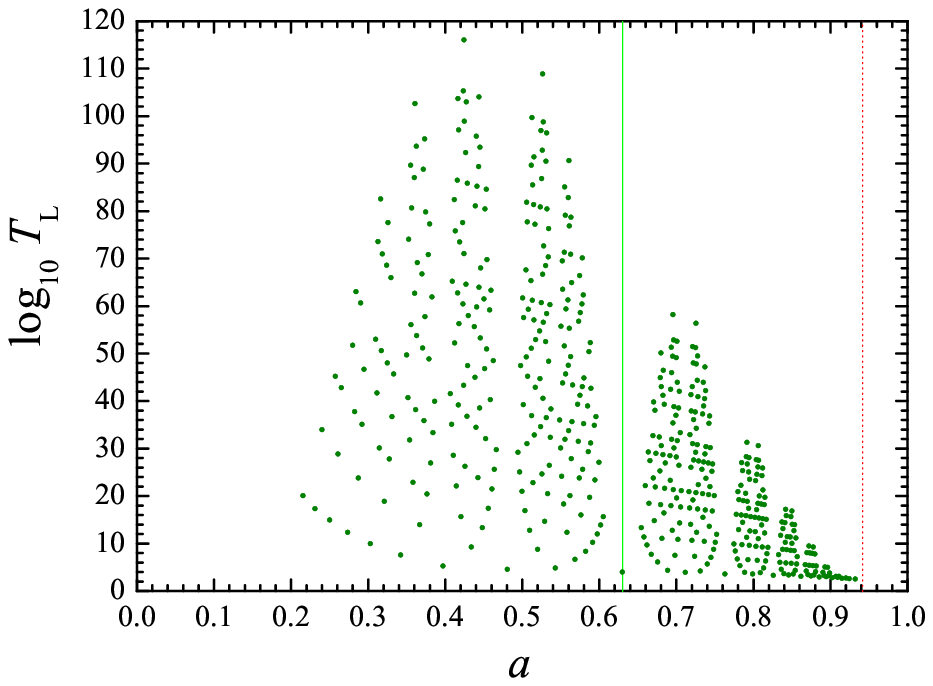}
&
\includegraphics[width=0.5\textwidth]{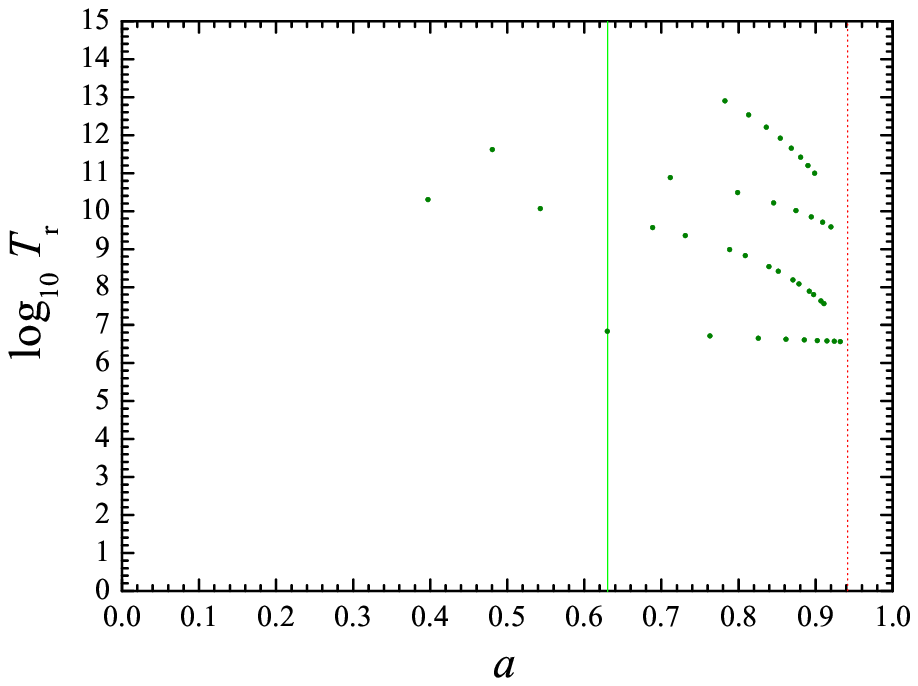}
\end{tabular}
\caption{The Lyapunov and removal times, in the same model as in
Fig.~\ref{SE_lgK_a}. Left panel: the Lyapunov time $T_\mathrm{L}$
(olive dots) as a function of the tertiary's semimajor axis $a$.
Vertical green line: the location of the 2/1 resonance with the
super-Earth. Vertical red line: the inner border of the
super-Earth's Wisdom gap. Right panel: the removal time
$T_\mathrm{r}$ (olive dots) as a function of semimajor axis $a$.
Solely the resonances with $K>K_\mathrm{G}$ are presented.}
\label{SE_TL_a}
\end{figure}

\begin{figure}[h!]
\centering
\includegraphics[width=0.8\textwidth]{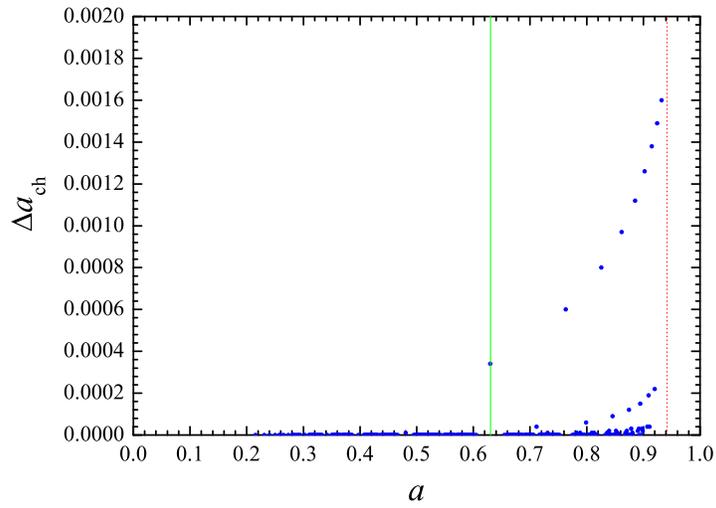} \\
\caption{The widths of subresonance multiplets of resonances, in
the same model as in Figs~\ref{SE_lgK_a} and \ref{SE_TL_a}. Blue
dots: the widths $\Delta a_\mathrm{ch}$ of subresonance
multiplets. For the resonances with $K < K_\mathrm{G}$ the widths
are set to zero. Green vertical line: the 2/1 resonance with the
super-Earth. Red vertical line: the inner border of the
super-Earth's Wisdom gap.} \label{SE_Width_a}
\end{figure}

\begin{figure}[h!]
\centering
\includegraphics[width=0.8\textwidth]{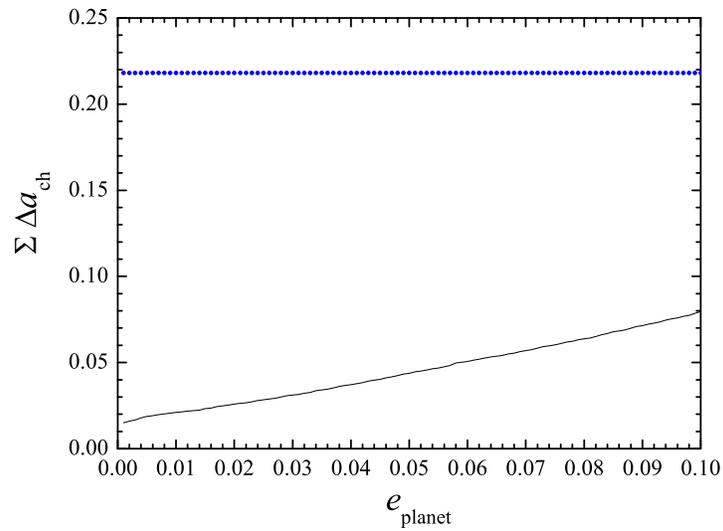} \\
\caption{The chaos covering factor in the inner Solar system (the
sum of the widths of the mean-motion resonances with $K >
K_\mathrm{G}$, up to the Farey tree level~10, as a function of
Jupiter's eccentricity $e_\mathrm{planet}$ (black solid curve).
The widths are in units of Jupiter's orbital radius. Blue dotted
horizontal line: the covering factor (the radial half-width) of
the Wisdom gap of Jupiter.} \label{W_ecc_mu-3}
\end{figure}

\begin{figure}[h!]
\centering
\includegraphics[width=0.8\textwidth]{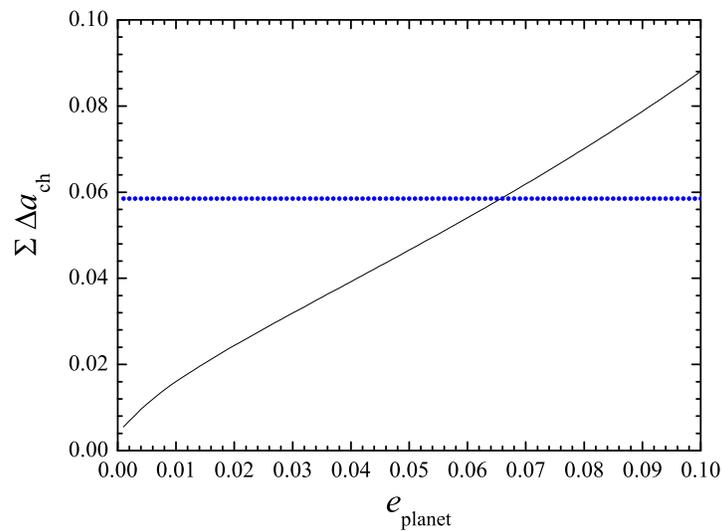} \\
\caption{The chaos covering factor in the inner model ``Sun-like
star -- super-Earth -- minor body'' system (the sum of the widths
of the mean-motion resonances with $K > K_\mathrm{G}$, up to the
Farey tree level~10), as a function of the super-Earth's
eccentricity $e_\mathrm{planet}$ (black solid curve). The widths
are in units of the super-Earth's orbital radius. Blue dotted
horizontal line: the covering factor (the radial half-width) of
the super-Earth's Wisdom gap.} \label{W_ecc_mu-5}
\end{figure}

Recall that the chaotic layers, if $\lambda$ is increased,
exponentially shrink in width \citep{C79,S08PLA}. On the other
hand, as it is clear from Figs~\ref{J_lgK_a}--\ref{J_TL_a} and
\ref{SE_lgK_a}--\ref{SE_TL_a}, the Farey tree of mean-motion
resonances forms two distinct major ``nests'', distinctly
separated from each other by the 2/1 resonance location at $\alpha
= \alpha_{1/2} =2^{-2/3} =0.630...$. The increase of $\lambda$
with decreasing $\alpha$ radically (exponentially with $\alpha \to
0$) suppresses chaos in the nest on the left of $\alpha_{1/2}$ (in
the panels of Figs~\ref{J_lgK_a}--\ref{SE_Width_a}), in contrast
to that on the right of $\alpha_{1/2}$, because the nests are far
from each other. In this way, the interplay of the rise of
$\lambda$ with decreasing $\alpha$ and the resonance nests' broad
separation explains the rather sharp EPCZ appearance.

The EPCZ phenomenon can be more or less active in determining the
architectures of planetary systems: the orbital zone between the
2/1 and 1/1 resonances with a planet can be expected to be
normally free from low-mass material and, perhaps, also from less
massive (than the perturber) planets. Only the material
occasionally captured in the first-order 3/2 or 4/3 resonance may
survive, as in the {\it Kepler}-223 system.

On the other hand, no restrictions apply to populate the zone
inner (in orbital radius) to the 2/1 resonance. In this respect,
the sharp difference in the global stability between the 0/1--2/1
and 2/1--1/1 orbital zones seems to agree with available data on
the known architectures of planetary systems.

This first of all concerns the observed structure of planetesimal
disks, such as the 2/1 resonance cut-off, in observed planetary
systems, including our Solar one. The main asteroid belt in the
Solar system is cut-off from above at its radial exterior by the
2/1 mean-motion resonance with Jupiter; therefore, if any material
have been ever substantially present in the 2/1--1/1 orbital zone
(corresponding to the radial space between the 2/1 and 1/1
mean-motion resonances with Jupiter), it has been exhausted,
whatever the reason for this removal could have been. Only some
small amount of material captured in the first-order 3/2 and 4/3
resonances could have survived. Note that the formation of
individual matter-free gaps at the 2/1 resonance is directly
observed in numerical experiments, already on relatively short
timescales \citep{DS16}.

Among observed exoplanet systems, a prominent example of the
2/1-resonance inner cut-off of a circumstellar disk is exhibited
in the HR~8799 system. This system is remarkable, being a
``young'' structural analogue of the Solar system. Indeed, its
architecture is similar to that of ours: the orbits of its
observed four giant planets are surrounded by a warm dust belt
analogous to the asteroid belt in our system, and from outside
they are surrounded by a cold belt analogous to the Kuiper belt
\citep{F21AJ}. The inner part of the system (bounded in radius
from above by the ``asteroid belt'') contains a zone of potential
habitability. According to \cite{F21AJ}, ``simply put, the system
of HR~8799 is a younger, broader, and more massive version of the
Solar System''. Its ``asteroid belt,'' in its turn, is cut-off
from above by the 2/1 resonance with the giant planet that is
innermost in this system, similar to the situation in the Solar
system.

What could be the mechanism responsible for any rapid-enough
removal of material from the weakly unstable zone? In our Solar
system, the Yarkovsky effect and the impact destruction (giving
birth to asteroid families) of bodies in the main asteroid belt
continuously supply material into numerous chaotic resonant bands
present inside the belt. This process monotonously, though slowly,
exhausts the belt: in the chaotic bands, the eccentricity is
slowly pumped up until the particles enter secular resonances, and
the latter drive the material away, mostly up to falling onto the
Sun (see, e.g., \citealt{Morbi02}). Note that the Yarkovsky drift
in the semimajor axis, $\mathrm{d}a/\mathrm{d}t$, can be estimated
using equations~(4)--(5) in \cite{BVRN06}. As illustrated in
Fig.~2 in \cite{BVRN06}, it may provide, depending on a number of
physical parameters, the rapid-enough permanent radial transport
of asteroidal material.

The same removal process is by all means active, to a more or less
degree, in planetesimal disks of any exoplanet system. Therefore,
it may more or less (depending on the system parameters) rapidly
exhaust the EPCZ. For this to occur, the ``clearing'' (those
providing the rapid-enough eccentricity pumping) chaotic resonant
bands should have a sufficient covering factor (as defined in
Section~\ref{sec_cfc}) in orbital radius.

In this article, we have considered the EPCZ formed interior to
the planet's orbit. Any global stability properties of the {\it
outer} resonance zones require a separate analysis, as they
broadly extend to infinity; it would be accomplished elsewhere. In
particular, this analysis could shed light on possible
resonant/chaotic structure of circumstellar external planetesimal
disks, similar to the Solar system's Kuiper belt, whose resonant
structure is mostly controlled by Neptune.

As follows from comparing the numerical results presented in
Sections~\ref{sec_SsSJ} and \ref{sec_msssE}, the dynamical
importance of the EPCZ in presence of smaller-$\mu$ perturbers
tends to be much greater than in presence of larger-$\mu$
perturbers: indeed, the results indicate that, for the Earths and
super-Earths orbiting the Solar-like stars, the removal of
material from their EPCZs is expected to be much more pronounced
than for the giant planets of similar host stars.

\medskip

\noindent {\bf Acknowledgments.} The author is most thankful to
the referee for comments and remarks. This work was supported in
part by the Russian Science Foundation, project 22-22-00046.

\medskip

\noindent {\bf Data availability.}
The data underlying this article will be shared on reasonable
request to the corresponding author.

\newpage

\end{document}